\documentstyle[preprint,aps,epsfig]{revtex}
\tightenlines
\begin{document}
\draft
\preprint{\vbox{
\hbox{IFT-P.056/99}
\hbox{July 1999}
}}
\title{Large magnetic dipole moments for neutrinos with arbitrary masses} 
\author{ J. C. Montero \footnote{e-mail: montero@ift.unesp.br}
and V. Pleitez\footnote{e-mail: vicente@ift.unesp.br}}  
\address{
Instituto de F\'\i sica Te\'orica\\
Universidade Estadual Paulista\\
Rua Pamplona 145\\ 
01405-900-- S\~ao Paulo, SP\\Brazil} 
\date{\today}
\maketitle
\begin{abstract}
We show that there is a general sort of models in which it is possible to 
have large magnetic dipole moments for neutrinos while keeping their masses 
arbitrarily small. Some examples of these models are considered.

\end{abstract}
\pacs{PACS numbers: 13.15.+g; 14.60.Pq; 14.60.St}


It has been known since several years ago that a way to solve the solar 
neutrino problem is provided by the assumption that the electron neutrino is 
a massive Dirac particle with a magnetic moment 
$\mu_\nu\leq(10^{-11}-10^{-10})\mu_B$, where $\mu_B$ is the Bohr 
magneton~\cite{ac,okun}. 
The bound on the muon neutrino magnetic moment, 
$\mu_{\nu_\mu}<10^{-8}\mu_B$, coming from 
neutral current data is larger than the one for the electron neutrino.
It has been pointed out that its effect can be already observed with the 
magnetic field of the Earth~\cite{kim1,kim2}. 
However, since there is a controversy over the validity of the claimed upper 
limits on the magnetic moment from the astrophysical data we use for reference
the values given in PDG~\cite{pdg}:
\begin{eqnarray}
&&\mu_{\nu_e}<3.2\times10^{-10}\mu_B,\nonumber \\&&
\mu_{\nu_\mu}<8.5\times10^{-10}\mu_B, \nonumber \\ &&
 \mu_{\nu_\tau}<5.4\times10^{-7}\mu_B.
\label{mmexp}
\end{eqnarray}

In the standard model its magnetic moment is rather small~\cite{fs} 
\begin{equation}
\mu_\nu\leq
(10^{-19}-10^{-18})(m_\nu/\mbox{eV})\mu_B.
\label{smmm}
\end{equation}

On the other hand, it is difficult to give to the neutrino a large magnetic 
moment because in most of models it is proportional to the neutrino mass.
In supersymmetric theories with $R$-parity breaking magnetic moments of
the order of $\mu_\nu\simeq 10^{-13}\mu_B$ arise~\cite{bgmt}.
This is worst in $L-R$ symmetric model in which 
$\mu_{\nu_e}<10^{-4}\mu_B$~\cite{bmr}.

The problem is that in most models the same loop diagram without the photon 
line give a contribution to the neutrino mass~\cite{bmr,zee}. 
To suppress the contribution to the mass
some authors introduce continuous~\cite{con} or discrete symmetries~\cite{dis}.
It is possible to have models in which an $SU(2)_\nu$ symmetry,
\`a la Voloshin~\cite{voloshin}, acting on $(\nu,\nu^c)$ as a doublet would
forbid the mass but allow the magnetic moment~\cite{bm}. It is possible
to implement a large $\mu_\nu$ in a model with $SU(2)\otimes U(1)$ gauge 
symmetry but with lepton doublets $(\nu\,l^-)_L$ and  
$(\nu\,l^-)_R$~\cite{kim3}.
The problems above arise because both the neutrino mass and
magnetic moment are supposed to be calculable. An economic model is a minimal
version of the Zee's model~\cite{zee} in which only one scalar singlet $h^-$
(besides the usual doublet) and right-handed neutrinos are added to the 
$SU(2)\otimes U(1)$ model~\cite{mf}.  

Here we will consider a general sort of models which have some of the 
feature of the models of Refs.~\cite{kim3,mf} in the sense that
the magnetic moment does not depend directly on the neutrino mass. A
similar mechanism for an electric dipole moment for the charged leptons
was proposed in Ref.~\cite{zee2}. In the present work we will
consider a more general sort of models with or without exotic charged leptons
and charged scalars. 

Here we will consider that the neutrino masses are in the same foot that the 
masses of all fermions: they are renormalizable and for this reason, they are 
not calculable. Finite or infinite contributions are erased by the 
renormalization procedure. On the other hand the magnetic moment of all 
elementary fermions are both finite and calculable.   

We will illustrate in this work the features of general models allowing
a neutrino magnetic moment which has this characteristic: the magnetic
moment is approximately independent of the neutrino mass since it appears
always as a factor $m_\nu/v_s$ where $v_s$ is a small vacuum expectation 
value (VEV) or it is in fact proportional only to the mass of a charged
antilepton (exotic or the usual ones). 
Two cases are going to be considered: {\it i)} all neutrinos are almost
degenerate in mass, so the condition $m_\nu/v_s\approx1$ is valid for all
of them; {\it ii)} there is a hierarchy in masses and the condition
$m_\nu/v_s\approx1$ is valid only for the heavier neutrino.

Supposing a model with right-handed neutrinos, we can parameterize the neutral 
and charged Higgs interactions in the lepton sector as follows:
\begin{eqnarray}
-{\cal L}^Y&=&\overline{E^\prime_L}\,G^EE^\prime_R\chi^0+
\overline{\nu^\prime}_L G^EE^\prime_R\chi^-\nonumber \\ &+&
\overline{\nu^\prime_L}G^\nu \nu^\prime_R\eta^0+
\overline{E^\prime_L}G^\nu\nu^\prime_R\eta^++H.c.
\label{1}
\end{eqnarray}
$G^E$ and $G^\nu$ are arbitrary complex matrices and assuming three neutrinos
$G^\nu$ is a $3\times 3$ matrix. 
Here $E^\prime$ can denote a positive charged exotic lepton $E^{\prime+}$ 
or the charge conjugated of the known charged lepton $l^{\prime+}$ and
the primes denote symmetry eigenstates (with respect to an arbitrary 
electroweak symmetry). 
With biunitary transformations like
\begin{equation}
\nu'_{L,R}={\cal O}^\nu_{L,R}\nu_{L,R},\quad
E'_{L,R}={\cal O}^E_{L,R}E_{L,R},
\label{2}
\end{equation}
with the unprimed field denoting mass eigenstates, we can redefine the 
interactions in Eq.~(\ref{1}) in terms of the mass matrices
$v_l{\cal O}^{E\dagger}_LG^E{\cal O}^E_R=M^E$, and
$v_s{\cal O}^{\nu\dagger}_LG^\nu{\cal O}^\nu_R=M^\nu$,
where $v_l=\langle \chi^0\rangle$ and $v_s=\langle \eta^0\rangle$ are
appropriate vacuum expectation values; 
$M^E$ and $M^\nu$ are real diagonal matrices, in particular
$M^E=\mbox{diag}(m_{E_1},m_{E_2},m_{E_3})$, $M^\nu=
\mbox{diag}(m_{\nu_1},m_{\nu_2},m_{\nu_3})$.

Next, we can rewrite the charged scalar interactions in Eq.~(\ref{1}) as
\begin{equation}
\frac{M^E}{v_l}\bar{\nu}_L K E_R\chi^-+
\frac{M^\nu}{v_s}\bar{E_L}K^\dagger \nu_R S^+
+H.c.,
\label{4}
\end{equation}
where $v_l$ and $v_s$ denote a large and a small vacuum expectation value 
(VEV), respectively, and $K= {\cal O}^{\nu\dagger}_L{\cal O}^E_L$. Moreover, 
$v_s$ is the only VEV which contributes to 
the neutrino masses. 
In some models the fraction $M^E/v_l$ and $M_\nu/v_s$ are substituted by 
dimensionless Yukawa couplings (see below). 

These interactions generate diagrams like the one shown in Fig.~1. 
Notice that one of the vertex is proportional to the neutrino mass, the other 
one is proportional to the mass of the charged lepton $E$ and there is still a
mass insertion of the charged lepton. Hence the magnetic moment is proportional
to $m_\nu m^2_E/v_\chi v_s$ times a dimensionless function.
 Explicitly we have that the magnetic moment for the $\nu_i$ neutrino, arisen 
from diagrams in Fig.~1 and the corresponding ones with the photon line 
attached to the internal fermion line, is given by 
\begin{equation}
\mu_{\nu_i}=-\frac{m_e}{ 4\pi^2} \frac{ m_{\nu_i}} { v_s }
 \sum_j \mbox{Re}\left( K^\dagger_{ij}K_{ji}\right)\frac{m_{E_j}}{v_l}
\frac{m_{E_j}}{m^2_{Y}}\;F(m_{Y},m_{E_j})\mu_B ,
\label{mmnus}
\end{equation}  
where $F=[F_+(m_{Y},m_{E_j})+F_-(m_{Y},m_{E_j})] $
and there is no sum in $i$; and we have also defined
\begin{eqnarray}
\lefteqn{F_\pm(m_{Y},m_{E_j})=
-\frac{m^2_{Y}}{2m^2_{\nu_i}}\ln\frac{m^2_{Y}}{m^2_{E_j} }+ 
\frac{m^2_{Y}}{2m^2_{\nu_i}\Delta} }
\nonumber \\ &\cdot& 
\left(m^2_{Y}\pm m^2_{\nu_i}-m^2_{E_j}\right)\,
\ln\left[ \frac{m^2_{E_j}+m^2_{Y}-m^2_{\nu_i}+\Delta}
{m^2_{E_j}+m^2_{Y}-m^2_{\nu_i}-\Delta}\right],
\label{fs}
\end{eqnarray}
with $\Delta^2=[(m_{Y}+m_{E_i})^2-m^2_{\nu_i}][(m_{Y}-m_{E_i})^2-
m^2_{\nu_i})]$,
where $Y^-$ is a mass eigenstate scalar but we have omitted the mixing
among $\chi^-$ and $\eta^-$ since this is a model dependent issue.
Notice that in Eq.~(\ref{mmnus}) we have already written $\mu_{\nu_i}$ in 
terms of the Bohr magneton $\mu_B=e/2m_e$. 

We can consider $m_{E_j}/v_l\approx1$ and the case {\it i)}
when the condition $m_{\nu_i}/v_s\approx1$ is valid for all neutrinos;
and in the limit $m_E,m_Y\gg m_\nu$ we can write 
\begin{eqnarray}
\mu_{\nu_i}&\approx& -\frac{m_e}{ 4\pi^2} 
 \sum_j \mbox{Re}\left( K^\dagger_{ij}K_{ji}\right)
m_{E_j}   \nonumber \\ &&\mbox{} 
\cdot \left[\frac{m^2_{Y}+m^2_{E_j}}{(m^2_{Y}-m^2_{E_j})^2 }\, 
\ln \left(\frac{m^2_{Y}}{m^2_E}\right)-\frac{2}{m^2_{Y}-m^2_{E_j} }
\right]\,\mu_B,
\label{mmnus2}
\end{eqnarray}
With ${\cal K}^\dagger_{e1}{\cal K}_{1e}
\approx1$, ${\cal K}^\dagger_{e 2}{\cal K}_{2e} \approx10^{-5}$ and
$ {\cal K}^\dagger_{e3}{\cal K}_{ 3e}\sim10^{-3}$ and $m_{E_1}\approx
m_Y$ and $m_{E_2,E_3}\not=m_Y$, 
we obtain values for the three $\mu$ compatible with the values given
in Eq.~(\ref{mmexp}). 
We see from Fig.~2 that for a given $j$ if $m_{E_j}\not=m_Y$ the respective 
$F$-factor contribution to Eq.~(\ref{mmnus}) is of the order of one.
However in Fig.~2 we do not include the $\sum_j \mbox{Re}
( K^\dagger_{ij}K_{ji})$ factor appearing in Eq.~(\ref{mmnus}). The later
factor is important for getting $\mu_\nu$ compatible 
with the constraints given in Eq.~(\ref{mmexp}).

For the case {\it ii)}, when the condition for the enhancement 
$m_\nu/v_s\approx1$ is valid only for the
heavier neutrino; $\mu$-values compatible with those in Eq.~(\ref{mmexp})
are also obtained but in this case there are suppression factors 
$m_{\nu_1}/m_{\nu_3}$ and $m_{\nu_2}/m_{\nu_3}$. 

\begin{figure}[ht]
\begin{center}
\centering\leavevmode
\epsfxsize=\hsize
\epsfbox{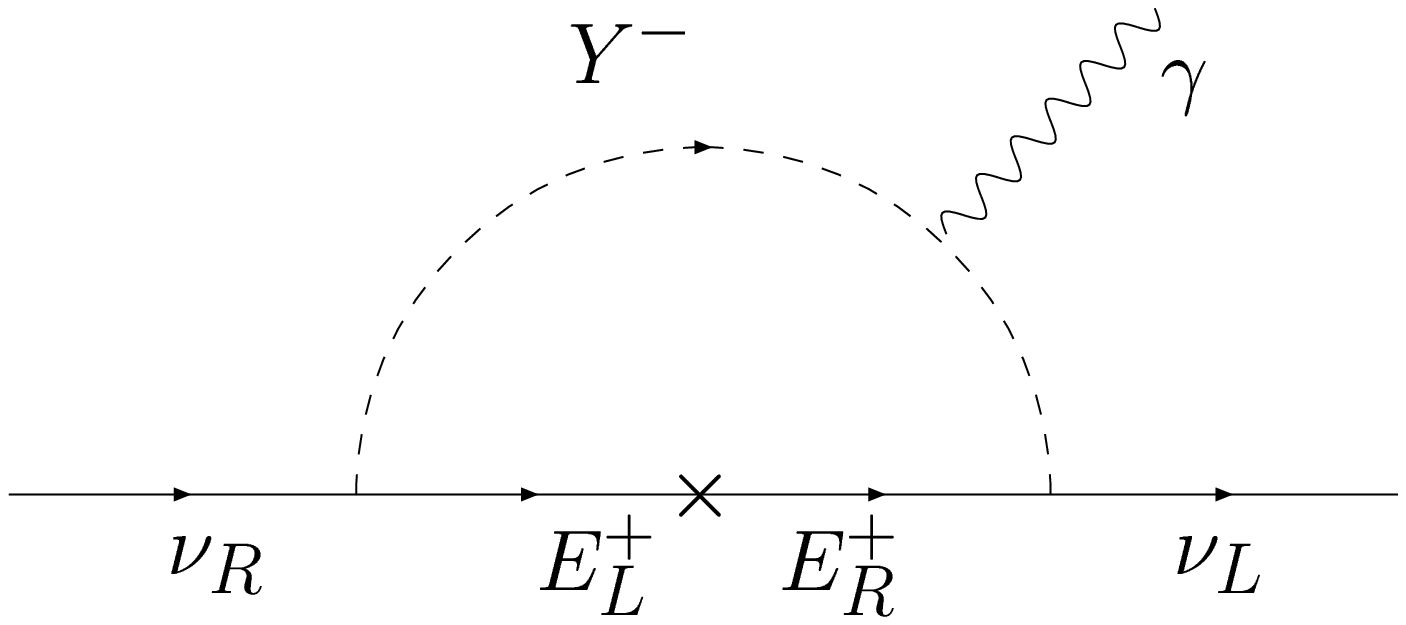}
\vglue -0.009cm
\end{center}
\caption{One loop contributions to the magnetic moment of the neutrinos.}
\label{fig1}
\end{figure}

Notice that the magnetic moments can be of the diagonal or transition type, 
hence for the $\nu_e$ case the phenomenological consideration of the 
resonant spin-flavor precession solution of the solar neutrino problem is 
valid~\cite{gn}.
So far all considerations are valid independently of the models. Any model
which contains interactions like those in Eq.~(\ref{1}) will produce a 
magnetic moment with the characteristic discussed above. For example, 
in multi-Higgs extension of the standard model, say with several doublets with
at least one of them coupling only to the leptons (by imposing an appropriate 
symmetry) plus a complex (non-majoron) triplet~\cite{cl}. In this case it is
possible to have FCNC in the charged lepton sector and the neutrino masses
are $m_\nu=G^\nu\,v_s$, where $v_s$ is the VEV of the  neutral component of the
triplet and $G^\nu$ is a complex symmetric $3\times 3$ matrix. 
There are also models
based on the $SU(3)_L\otimes U(1)_N$ electroweak symmetry with {\it a)} 
the leptons in triplets $\psi=(\nu_l,l^-,E^+_l)^T$~\cite{pt} or, {\it b)} 
$\psi=(\nu_l,l^-,l^+)^T,\;l=e,\mu,\tau$~\cite{pp}. In both models we 
have to add right-handed neutrinos. 
In the first model it is necessary also to add a scalar sextet $S$
which is not needed in the minimal version of the model and we denote the
VEV which give a contribution to the neutrino mass as $v_1$.
(The other neutral component of the sextet can give  
contributions to the charged lepton masses.) 
In this situation, the Yukawa interactions are 
\begin{equation}
-{\cal L}_{l}=\frac{G_{ab}}{\sqrt2}\,
\overline{(\psi_{aiL})^c}\,\psi_{bjL}S_{ij}+
\frac{1}{2} G'_{ab}\overline{(\psi_{aiL})^c}\,\nu_{bR}\eta
+H.c.,
\label{yuka}
\end{equation}
and the mass matrix of the neutrinos are of the form
\begin{equation}
\left(\begin{array}{cc}
Gv_s & \frac{1}{2}G'v_\eta\\
\frac{1}{2}G'v_\eta & M^R
\end{array}
\right)
\label{nmm}
\end{equation}
where $M^R$ is a possible Majorana mass term for the right-handed singlets
which we have not included in Eq.~(\ref{yuka}).
Hence, the biunitary matrices which diagonalize the mass matrix in 
Eq.~(\ref{nmm}) do not diagonalize separately neither $Gv_S$ nor $G'v_\eta$, 
and there are flavor changing neutral interactions in the lepton-Yukawa sector.
The interactions with the charged scalar are like those in Eq.~(\ref{1})
where $E_j$ denote  exotic charged leptons~\cite{pt} in the {\it a)} case, or 
the usual antileptons $e^+,\mu^+,\tau^+$~\cite{pp} in the {\it b)} one. 
Notice that now in Eq.~(\ref{4}) we have used $M^\nu/v_s\to G'\approx 1$. 
If the neutral interactions of the leptons $E_j$ also violate flavor,
both vertices in the diagram in Fig.~\ref{fig1} are not proportional
to the lepton masses and the model is of the type of the model of 
Ref.~\cite{zee2}. The same happens with the first vertex (from the left) if 
the exotic leptons $E_j$ have mixing with the known leptons. 
This is the case of the 331 model of Ref.~\cite{pp}. 
In both models neutrinos are Majorana particles and may still have transition 
magnetic moments. However the general sort of models which are parameterized
like in Eqs.~(\ref{1}) or (\ref{4}), neutrinos can be Dirac particles with
diagonal and transition dipole magnetic moments.
In models with exotic charged leptons there is no contribution to
the $\mu\to e\gamma$ decay if the exotic leptons do not couple to the usual
known leptons as in Ref.~\cite{pt}. In that model the contribution to the 
$\mu \to e\gamma$ decay arise via doubly charged scalars. 
On the other hand in models with only the known leptons the constraints
coming from the decay $\mu\to e\gamma$ are~\cite{mf}
\begin{equation}
\frac{{\cal K}^\dagger_{e 2}{\cal K}_{2e} }{m^2_Y}<10^{-8}\,\mbox{GeV}^{-2},
\label{cueg}
\end{equation}
which constrains only the $\mu_{\nu_\mu}$.

Of course, neutrinos have to have  
a mass different from zero for having $\mu_\nu\not=0$ but, as we said before,
that mass is arbitrary and could be rather small.
We have been able to give a mechanism to generate a $\mu_\nu$ which
can be in the range $10^{-13}$--$10^{-11}\mu_B$ {\it i.e.}, as large as the 
current upper limit coming from the supernovae collapse~\cite{mex}
\begin{equation}
\mu_{\nu_e}<(0.1-0.4)\times 10^{-11}\mu_B,
\label{muexp}
\end{equation}
or those in Eq.~(\ref{mmexp}), even for a neutrino with an eV mass or 
smaller. However if their masses are small the respective mass 
square differences must also be very small and could be compatible with
all experimental data coming from neutrino experiments like 
accelerator~\cite{lsnd}, solar~\cite{solarexp}, and 
atmospheric~\cite{sk} ones. 
 
Up to now the analyses of atmospheric neutrinos were restricted by events
induced by the charged currents $(CC)$ interactions ($e$-like and $\mu$-like 
events). However, events induced by neutral currents $(NC)$ can give important
information on the neutrino flavor oscillations~\cite{pi0}. For example,
a precise measurement of the ratio of the $\pi^0$-like events to the $e$-like
events~\cite{sk2} can be used to distinguish $\nu\to\nu_\tau$ from
$\nu_\mu\to \nu_s\,\mbox{sterile  neutrino}$ oscillation. However, if there 
exist a large muon neutrino magnetic moment (diagonal or transition), it will 
produce an additional
neutral current effect which has to be separated out to draw a definite 
conclusion~\cite{kim4}.  
We have  seen that it is possible to have
neutrinos with a magnetic dipole moment as large as $(10^{-11}-
10^{-10})\mu_B$ even with masses compatible with the mass square differences 
needed in LSND~\cite{lsnd}, solar~\cite{solarexp} and atmospheric~\cite{sk} 
neutrinos data.

Finally, notice that in Eq.~(\ref{mmnus}) if $K$ is a general unitary matrix, 
the interactions in Eq.~(\ref{1}) will induce electric dipole moments too.

\acknowledgments 
This work was supported by Funda\c{c}\~ao de Amparo \`a Pesquisa
do Estado de S\~ao Paulo (FAPESP), Conselho Nacional de 
Ci\^encia e Tecnologia (CNPq) and by Programa de Apoio a
N\'ucleos de Excel\^encia (PRONEX).

\vglue 0.01cm

\newpage

\vglue 0.01cm
\begin{figure}[ht]
\begin{center}
\centering\leavevmode
\epsfxsize=400pt
\epsfbox{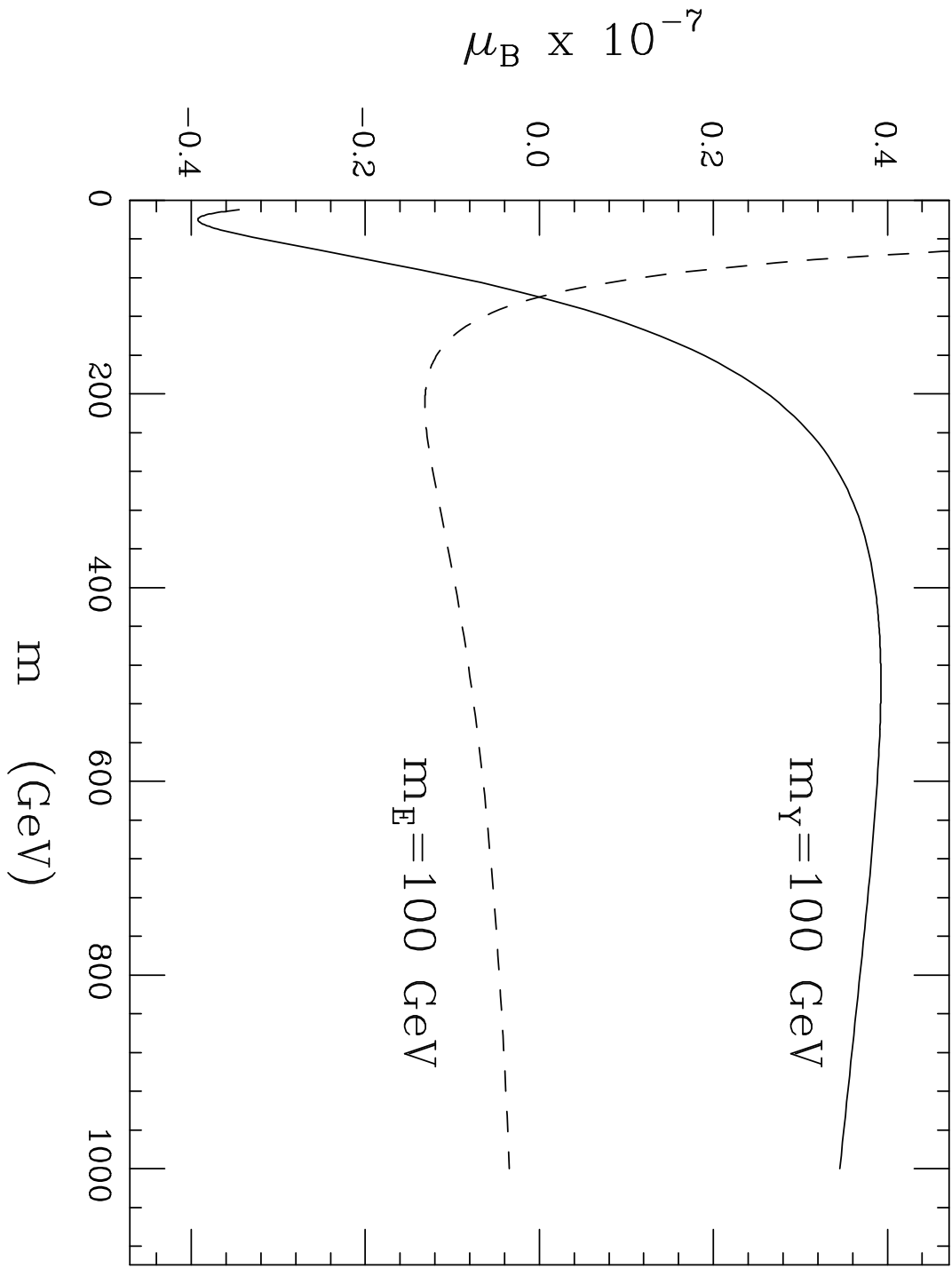}
\vglue -0.009cm
\end{center}
\vglue 2cm
\caption{ The neutrino magnetic moment from Eq.~(7) up to the mixing factor and
for a fixed $j$ and $m_Y$ ($m_E$) as a function of $m_E$ ($m_Y$).}
\label{fig2}
\end{figure}

\end{document}